# Control of ultrafast pulses in hydrogen-filled hollow-core photonic crystal fiber by Raman coherence

F. Belli[1,*], A. Abdolvand[1], J. C. Travers[1,2] and P. St.J. Russell[1]

[1]Max Planck Institute for the Science of Light, Staudtstrasse 2, 91058 Erlangen, Germany
[2]School of Engineering and Physical Sciences, Heriot-Watt University, Edinburgh, EH14 4AS, UK

We present the results of an experimental and numerical investigation into temporally non-local coherent interactions between ultrashort pulses, mediated by Raman coherence, in gas-filled kagomé-style hollow-core photonic crystal fiber. A pump pulse first set up the Raman coherence, creating a refractive index grating in the gas that travels at the group velocity of the pump pulse. Varying the arrival time of a second probe pulse allows high degree of control over its evolution as it propagates along the fiber, in particular soliton self-compression and dispersive wave (DW) emission. In the experiments reported, a DW is emitted at ~300 nm, with a central frequency that oscillates with the pump-probe delay. The results demonstrate that strong Raman coherence created in broadband guiding gas-filled kagomé-PCF can be used to control the dynamics of ultrashort probe pulses, even in difficult-to-access spectral regions such as the deep and vacuum ultraviolet.

*Introduction*—Controlling the nonlinear propagation of ultrashort pulses in gases is of high interest in many fields, for example high harmonic generation [1] and spectroscopy [2,3]. Gas-filled hollow-core photonic crystal fiber (HC-PCF) has proven in recent years to be an ideal vehicle for enhancing all kinds of gas-laser interactions [4,5]. Kagomé-style HC-PCF ("kagomé-PCF") in particular has been used to generate bright ultraviolet (UV) and vacuum UV (VUV) dispersive waves from few-µJ ~50 fs pump pulses in the near infrared [6-8]. Such sources are readily scalable to high average powers using compact high repetition rate fiber lasers [9]. Controlling the characteristics of these ultrashort UV-VUV pulses [10,11] is extremely challenging, often requiring the development of new techniques. Here we report a pump-probe technique that makes use of the long-lived Raman coherence [12,13] ("memory" [14]) in hydrogen-filled kagomé-PCF. At gas pressures of a few atmospheres, the coherence time $T_2$ of the molecular oscillations is of order several 100 ps, which is more than 1000 times longer than the pump and probe pulses. In the experiments, a pump pulse first creates a Raman coherence wave and then a probe pulse, arriving within time $T_2$, is modulated by it. The technique has similarities with the coherent control [15] of plasma dynamics [16] and multiphoton transitions [17].

*Physical model*—The propagation of ultrashort pulses is accurately described by the unidirectional pulse propagation equation (UPPE) [18,19], as confirmed both in free-space configurations [20] and in waveguide geometries such as hollow dielectric capillaries and PCF [6,7,11,19]. The UPPE can be written, for the $LP_{01}$-like core mode, in the form:

$$\partial_z \tilde{E}_1(z,\omega) = i\left[\beta(\omega) - \frac{\omega}{v_1}\right]\tilde{E}_1(z,\omega) + i\frac{\mu_0 \omega^2}{2\beta(\omega)}\tilde{p}_{11}(z,\omega) \quad (1)$$

where $z$ is the position along the fibre, $\omega$ the angular frequency, a tilde above a quantity denotes the temporal Fourier transform, and $E(z,t)$ is the electric field, defined over an effective mode area $A_{eff}$. The modal wavevector is $\beta(\omega)$, $c$ is the speed of light in vacuum, $\epsilon_0$ the permittivity of free space, $p_{11}(z, \omega)$ the nonlinear polarization, $v_1$ the group velocity of a pulse with central frequency at $\omega_0$ and $\tau = t - z/v_1$ is the reduced time in a reference frame moving at velocity $v_1$.

In order to isolate inter-pulse effects, we consider two highly delayed pulses:

$$\begin{aligned}E_1(0,t) &= \sqrt{2P_1/(n_{eff}\varepsilon_0 cA_{eff})}\, e^{-t^2/2\tau_p^2}\cos(\omega_0 t) \\ E_2(0,t) &= \sqrt{2P_2/(n_{eff}\varepsilon_0 cA_{eff})}\, \exp^{-(t-\tau_{in})^2/2\tau_p^2}\cos(\omega_0 t)\end{aligned} \quad (2)$$

where 1 represents the pump pulse and 2 the delayed pulse. $P_i$ is the peak power of the $i$-th pulse, $\tau_p$ its duration, and $n_{eff}$ is the effective modal index. The delay $\tau_{in}$ between pulses is defined so that $E_1(z,t)E_2(z,t)$ is effectively zero, i.e., $\tau_{in} \gg \tau_p$. In this highly non-overlapping limit, the first pulse is described by Eq. (1), but the second pulse obeys the equation:

$$\partial_z \tilde{E}_2 = i\left[\beta(\omega) - \frac{\omega}{v_1}\right]\tilde{E}_2 + i\frac{\mu_0 \omega^2}{2\beta(\omega)}\left[\tilde{p}_{12} + \tilde{p}_{22}\right]. \quad (3)$$

In Eq. (1) and Eq. (3), $p_{11}$ and $p_{22}$ account for any *self-induced* nonlinear polarization waves created by the first and second pulses. This may include instantaneous and non-instantaneous nonlinear responses such as Kerr-induced self-phase modulation and Raman- and ionization-induced nonlinear polarization. On the other hand, $p_{12}$ accounts only for any causal and long-lived (compared to the pump pulse duration) nonlinear polarization acting on the probe pulse. For example, the long-lived Raman coherence or the free-electron density produced by the first pulse may survive to influence the second pulse via the term $p_{12}$. Note that the Kerr-induced cross-phase modulation ($\propto E_1^2 E_2$ [21]) of the probe pulse does not play any role in $p_{12}$ in Eq. (2), since the two pulses do not overlap in time.

Here we focus our attention on the effect of the long-lived Raman coherence wave, excited impulsively [6,12,13,22] by the pump pulse, on the probe pulse for delays $\tau \gg \tau_p$ but within the lifetime of the Raman coherence wave $T_2$, i.e., $\tau \ll T_2$. Under these circumstances $p_{12}(z, \tau)$ can be written:

$$p_{12}(z, \tau_p \ll \tau \ll T_2) = 2n_0 \varepsilon_0 \Delta n_{12}(z,\tau) E_2(z,\tau)$$
$$\Delta n_{12}(z,\tau) \simeq \frac{N_t \alpha_R^2}{2n_0 \varepsilon_0 \hbar} \left| \tilde{I}(\Omega_R, z) \right| \sin(\Omega_R \tau) \quad (4)$$

where $N_t$ is the total molecular density and $\Delta n_{12}(z, \tau)$ is the resulting nonlinear refractive index modulation, which is proportional to the $z$-dependent intensity spectrum of the pump pulse at the Raman frequency $\Omega_R/2\pi$:

$$\tilde{I}(\Omega_R, z) = \int_{-\infty}^{+\infty} dt E^2(z,t) e^{i\Omega_R t}$$
$$= \frac{1}{2\pi} \int_{-\infty}^{+\infty} d\omega' \tilde{E}(\omega', z) \tilde{E}^*(\omega' - \Omega_R, z). \quad (5)$$

Figure 1(a) shows the resulting nonlinear refractive index modulation $\Delta n_{12}(0, \tau)$ when the system is driven in the impulsive regime, i.e., $\tau_p \ll T_R = 2\pi/\Omega_R$ [22]. In contrast to glass, where the Raman coherence time $T_2$ ranges from a few fs in fused silica to ~250 fs in As$_2$S$_3$, $T_2$ in molecular gases ranges from hundreds of ps to a few ns, depending on the gas pressure, making pump-probe measurements with fs-pulses rather straightforward to carry out. Fig. 1 illustrates such a pump-probe experiment (numerically modeled) with identical pump and probe pulses as the delay of the probe is scanned over $T_R$. Figs. 1(b-e) show the total nonlinear refractive index modulation $\Delta n_{tot} = \Delta n_{22} + \Delta n_{12}$ (black curves) for the four different values of $\tau_{in}$ marked by black dots in Fig. 1(a). $\Delta n_{tot}$ carries information on the Raman-induced refractive index change induced both by the first ($\Delta n_{12}$) and the second pulses ($\Delta n_{22}$). When $\tau_{in} = mT_R$, where $m$ is an integer, the two Raman responses add (with relative phase $\phi = 0$), so that the probe pulse sees an increasing refractive index in the vicinity of $\tau = \tau_{in}$. When $\tau_{in} = (m + \frac{1}{2})T_R$, however, the Raman responses are $\pi$ out-of-phase (Fig. 1(d)) and the probe pulse sees a decreasing refractive index under its envelope. When $\tau_{in} = (m + \frac{1}{4})T_R$ the overall Raman response is bell-shaped (Fig. 1(c)), with regions of increasing and decreasing index under the probe pulse envelope.

These refractive index variations strongly affect the dynamics of the probe pulse as it propagates along the fiber (Fig. 2(a-d)). For $\phi = 0$ it is red-shifted, with a steepened trailing edge (Fig. 2(a)). For $\phi = \pi$ there is an overall blue-shift, with a steepened leading edge (Fig. 2(b)), and for $\phi = \pm \pi/2$ there is either pulse compression (Fig. 2(c)) or decompression (Fig. 2(d)).

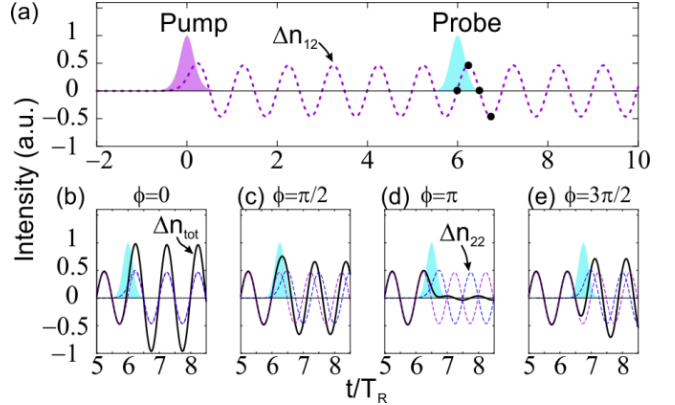

Fig. 1: (Color online). (a) Pump (purple) and probe (cyan) intensities together with nonlinear modulation of the refractive index induced by the pump $\Delta n_{12}$ (dashed purple) as a function of the normalized temporal delay $\tau/T_R$. (b-d) Probe intensity and self-induced Raman nonlinear index induced by the probe (dashed blue, $\Delta n_{22}$) and by the pump (dashed purple, $\Delta n_{12}$), altogether with overall Raman nonlinear index $\Delta n_{tot} = \Delta n_{22} + \Delta n_{12}$.

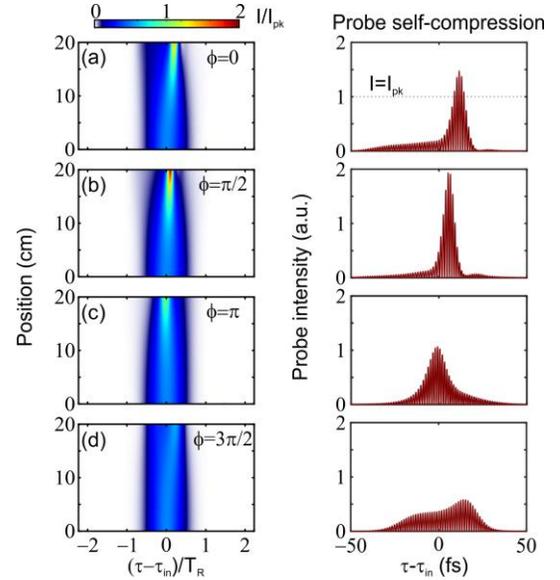

Fig. 2: (Color online) Left: Numerically simulated probe intensity as a function of time $\tau/T_R$ and fiber position for $\tau_{in}/T_R = 6.5, 7, 7.5$ and 8. Right: Snapshots of the probe intensity profile at the fiber output (red lines). All the probe intensities are normalized to the peak pump intensity $I_{pk}$.

*Controlling self-compression of the probe pulse*—In the experiment, a pulse shaper (Dazzler) at the fiber input was used to create, and control the delay between, two identical, transform-limited (35 fs) pulses, each with 0.2 μJ energy.

These were delivered from a 1 kHz Ti:sapphire laser system with central wavelength at 805 nm. The inter-pulse delay $\tau_{in}$ and pulse shapes at the fiber input were estimated from measured FROG traces. A kagomé-PCF, with a flat-to-flat core diameter of 25.3 μm and length of 20 cm, was filled with 7 bar of hydrogen. For these parameters the zero dispersion wavelength is 498 nm, the second order dispersion is –3.2 fs$^2$/cm, and the pulses undergo moderate soliton self-compression [4,5] with soliton order ~2.

Fig. 3(a) shows the experimental spectra collected at the fiber output as function of the input delay $\tau_{in}$. The spectral fringes observed at each value of $\tau_{in}$ arise from the interference of the two pulses in the spectrometer [23,24]. Periodic breathing of the spectral bandwidth, which occurs every $T_R$ = 57 fs (the period of the S(1) transition in ortho-hydrogen [25] and marked by vertical lines) is evident as well (note that the duration of the pump and probe pulses is not short enough to excite vibrational transitions in H$_2$ ($T_R$ = 8 fs) as in [6]). The spectral bandwidth reaches its maximum extent for $\tau_{in} = (m + ¾)T_R$, as expected from the numerical simulations in Fig. 2(c). There is excellent agreement, without any free parameters, between experiment and numerical modeling, as seen in Fig. 3(b).

The spectral modulation at fixed $\tau_{in}$ is caused by spectral interference of the pump and probe fields in the spectrometer, and takes the form [22]:

$$\tilde{S}(\omega;\tau_{in}) = |\tilde{E}_1(\omega)|^2 + |\tilde{E}_2(\omega)|^2 \\ + 2|\tilde{E}_1(\omega)||\tilde{E}_2(\omega)|\cos(\varphi_1(\omega)-\varphi_2(\omega)) \quad (6)$$

where $\varphi_1(\omega)$ and $\varphi_2(\omega)$ are the spectral phases of the pump and probe pulses at the fiber output. Eq. (6) encodes information on the phase difference between the two pulses. The spectral fringe spacing in Fig. 3(a) is determined by their separation $\tau_{out}$ at fiber output. As result of the Raman-related change in the refractive index (Eq.(4)), $\tau_{out}$ "wiggles" around $\tau_{in}$ with a period that equals $T_R$, as we will now discuss.

In order to extract information on the relative phase from the observed spectral fringes, and thus the temporal separation between the pulses at the fiber output, we take the inverse Fourier transform of $\tilde{S}(\omega;\tau_{in})$. The resulting function includes the sum of the autocorrelation functions of the pump and the probe pulses, $I_{ac}(\tau;\tau_{in}) = E_1 \otimes E_1 + E_2 \otimes E_2$, and twice their cross-correlation $I_{xc}(\tau;\tau_{in}) = E_1 \otimes E_2$, where $\otimes$ represents the convolution and $\tau$ is the pulse separation at the fiber output, estimated from the fringe spacing in Fig. 3(a). Fig. 3(c) plots $S(\tau;\tau_{in}) = I_{ac}(\tau;\tau_{in}) + 2I_{xc}(\tau;\tau_{in})$, calculated from the experimental data in Fig. 3(a). Since the cross- and auto-correlations are even functions of $\tau$, we plot only the $\tau > 0$ part. It consists of a contribution at $\sim\tau_p/2$ and two contributions centered at $|\tau| \sim \tau_{in}$. $I_{ac}(\tau;\tau_{in})$ maps the periodic change in probe pulse duration, which is inversely related to the spectral bandwidth breathing observed in Fig. 3(a). On the other hand, $I_{xc}(\tau;\tau_{in})$ maps the relative phase difference, and thus the temporal spacing, at the fiber output. The black dashed line in Fig. 3(c) is for $\tau = \tau_{in}$. It is evident that $I_{xc}$ has a peak at $|\tau| \sim \tau_{in}$. Defining the position of the peak as the actual temporal separation $\tau_{out}$ between the two pulses at the fiber output (i.e., the reciprocal of the spectral fringe spacing), it is clear from Fig. 3(c) that $\tau_{out}$ oscillates around $\tau_{in}$. Note that the increase in probe bandwidth is commensurate with its reduced pulse duration.

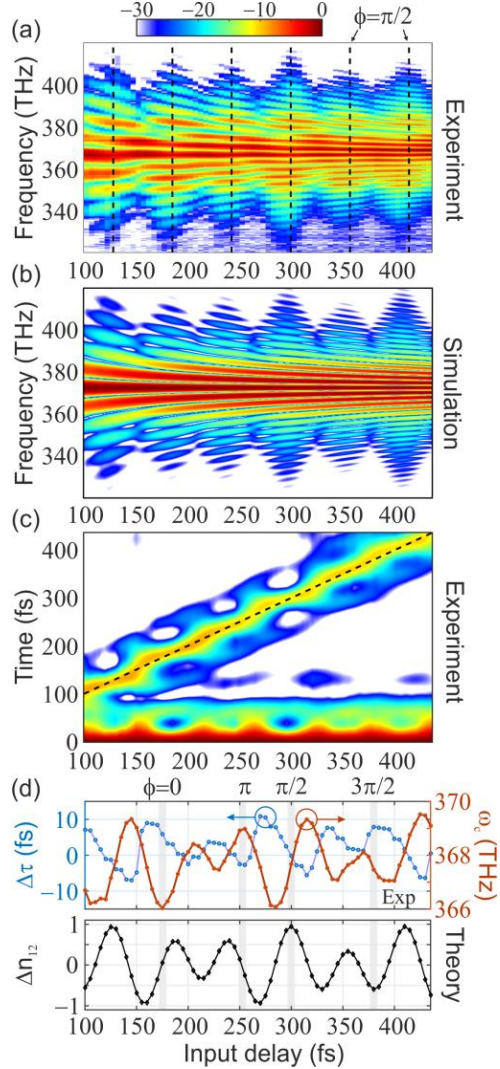

Fig. 3: (Color online) (a) Experimental spectra at the fiber output as function of $\tau_{in}$. (b) Corresponding numerical simulations. (c) Cross-correlation function $I_{xc}(\tau; \tau_{in})$ as a function of time and $\tau_{in}$, obtained from the experimental spectra in (a). (d) Lower: Theoretically calculated Raman-related refractive index modulation. Upper: calculated central frequency $\omega_c$ (red curve, right-hand axis) and temporal delay $\Delta\tau = \tau_{out} - \tau_{in}$ (left-hand axis, a.u.).

Fig. 3(d) plots $\Delta\tau = \tau_{out} - \tau_{in}$ versus $\tau_{in}$, highlighting the acceleration ($\Delta\tau < 0$) and deceleration ($\Delta\tau > 0$) experienced by the probe pulse during propagation [26] and caused by pump-induced rotational coherence. Since the pump pulse is short enough to excite rotational Raman coherence in both para- and ortho-hydrogen, both contributions must be taken into account for a complete explanation of the observed dynamics. Fourier analysis of $\Delta\tau(\tau_{in})$ reveals two strong sidebands centered at 10±1 THz and 17±1 THz, which match the rotational S(1)-transitions of ortho- and para-hydrogen, centered at 17.6 and 10.8 THz respectively. Fig. 3(d) shows (black line, lower panel) the expected overall modulation of nonlinear refractive index due to the cross-term:

$$\Delta n_{12}(z,\tau) = \frac{N_t \alpha_R^2}{2 n_0 \varepsilon_0 \hbar} \Big[ w_{para} \tilde{I}(\Omega_{para}, z) \sin(\Omega_{para} \tau) \\ + w_{ortho} \tilde{I}(\Omega_{ortho}, z) \sin(\Omega_{ortho} \tau) \Big] \quad (6)$$

where $w_{para} = 0.33$ and $w_{ortho} = 0.67$ are the relative populations of para- and ortho-hydrogen under standard conditions. The oscillation of the mean central frequency $\omega_c(\tau_{in})$, obtained from the spectra in Fig. 3(a) and plotted as the orange curve in Fig. 3(d), fits well to the value estimated from Eq. (6) (Fig. 3(d), black curve).

From the experimental curves we observe an increase in the red-shift and probe deceleration for $\phi = 0$, while for $\phi = \pi$ we observe a blue-shift and probe acceleration. The maximum spectral bandwidth and minimum temporal duration of the $I_{ac}$ contribution is observed for $\phi = \pi$, and the minimum bandwidth and largest duration for $\phi = \pi/2$, in complete agreement with the numerical simulations shown in Fig. 2(a-d).

*Dispersive wave wiggling*—We now discuss inter-pulse effects that occur when the input pulse undergoes strong soliton self-compression, resulting in the generation of a supercontinuum (SC) and emission of a dispersive wave. The experimental parameters were chosen so that both pulses had soliton order ~ 5. In contrast to the previous case, however, the pump pulse carried slightly less energy so as to avoid pulse break-up and dispersive wave (DW) emission, with the result that the observed DW dynamics originated entirely from the delayed probe pulse. Using a dispersion-balanced Mach-Zehnder interferometer as pulse shaper, we increased the temporal resolution to 1 fs and widened the range of the inter-pulse delay to 3 ps. The resulting spectra, obtained in a 20 cm length of kagomé-PCF with a 28 µm core diameter and filled with 15 bar of hydrogen, are plotted as a function of input delay in Fig. 4(a). Both pulses are well within the anomalous dispersion regime, and the pump pulse with energy 0.5 µJ, and the probe pulse 0.6 µJ. The pressure was chosen so as to simplify the detection of the emitted DW, which is at ~300 nm, well within the spectral range (200 to 1100 nm) of the CCD-based spectrometer. The fiber length and pulse energies were also optimized experimentally to obtain DW emission at a position as close as possible to the fiber end, so as to avoid any further nonlinear evolution of the DW [6]. A complex periodic modulation of the spectra is observed as function of inter-pulse delay, spanning the whole spectrum (300 to 1100 nm). There is a periodic enhancement of the spectral power at longer wavelengths (800 to 1100 nm) and a corresponding oscillation of the central frequency of the ~300 nm DW band. A SC spectrum, flat to better than 20 dB, is clearly observed every 57 fs (ortho-hydrogen), as shown in Fig. 4.

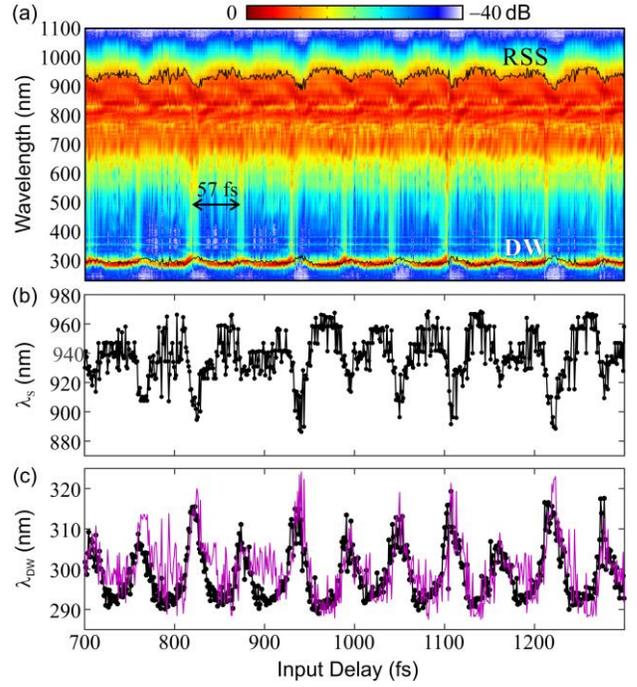

Fig. 4: (Color online) (a) Experimental spectra collected at the fiber output, plotted against the pump-probe delay $\tau_{in}$ at the fiber input. (b) The wavelength $\lambda_s$ of the strongest red-shifted part of the spectrum (RSS). (c) The central wavelength $\lambda_{DW}$ of the UV dispersive wave. The purple line is the DW wavelength, calculated assuming $\lambda_s$ is the central wavelength of the compressed probe pulse.

To understand these dynamics, we plot the delay dependence of the wavelength $\lambda_s$ of the strongest contribution at wavelengths longer than 800 nm (Fig. 4(a)), and the mean wavelength of the emitted DW (Fig. 4(b)). Taking $\lambda_s$ to be the effective central wavelength of the compressed probe pulse, the wavelength of the emitted DW can be estimated from phase-matching (purple curve in Fig. 4b) [5]. The more the compressed pulse is decelerated (red-shifted) the larger is the acceleration (blue-shift) of the DW, which empirically explains the anti-correlation observed between $\lambda_s$ and $\lambda_{DW}$ in Fig. 4. As $\tau_{in}$ is varied, the redshift and blue-shift oscillate with a period given by the $2\pi/\Omega_R$. We remark that this wiggling of the DW frequency is a general phenomenon that will occur in many different experiments, including those involving the break-up of a single pulse [6].

*Conclusions*—the nonlinear dynamics of a probe pulse, including supercontinuum generation and dispersive wave emission, can be simply controlled by varying its time delay relative to a pump pulse that creates coherent Raman oscillations in hydrogen-filled kagomé-PCF. Excellent agreement is obtained between experiment, analytical theory and numerical simulations. This novel system offers a versatile highly non-instantaneous means of controlling of the dynamics of ultrashort pulses in gas-filled PCF, for example, in pulse-compression, spectral broadening of DUV light and vacuum-UV emission.

* f.belli@hw.ac.uk